\documentclass[aps,pra,twocolumn,superscriptaddress]{revtex4-2}
\usepackage{graphicx}
\usepackage{amsmath}
\usepackage{amssymb}
\usepackage{color}
\usepackage{physics}
\begin{document}
\title{Promoting quantum correlations in DQC1 model via post-selection}

\author{Elisa I. Goettems}

\affiliation{Departamento de F\' \i sica, Universidade Federal de Santa Catarina, Florian\'opolis, SC, 88040-900, Brazil}

\author{Thiago O. Maciel}

\affiliation{Departamento de F\' \i sica, Universidade Federal de Santa Catarina, Florian\'opolis, SC, 88040-900, Brazil}

\author{Diogo O. Soares-Pinto}

\affiliation{Instituto de F\'isica de S\~ao Carlos, Universidade de S\~ao Paulo CP 369, 13560-970, S\~ao Carlos, S\~ao Paulo, Brazil}

\author{E. I. Duzzioni}

\affiliation{Departamento de F\' \i sica, Universidade Federal de Santa Catarina, Florian\'opolis, SC, 88040-900, Brazil}


\date{\today}
\begin{abstract}
The deterministic quantum computation with one qubit (DQC1) model is a restricted model of quantum computing able to calculate efficiently the normalized trace of a unitary matrix. In this work we analyse the quantum correlations named entanglement, Bell's nonlocality, quantum discord, and coherence generated by the DQC1 circuit considering only two qubits (auxiliary and control). For the standard DQC1 model only quantum discord and coherence appear. By introducing a filter in the circuit we purify the auxiliary qubit taking it out from the totally mixed state and consequently promoting other quantum correlations between the qubits, such as entanglement and Bell's nonlocality. Through the optimization of the purification process we conclude that even a small purification is enough to generate entanglement and Bell's nonlocality. We obtain, in average, that applying the purification process repeatedly by twelve times the auxiliary qubit becomes $99\%$ pure. In this situation, almost maximally entangled states are achieved, which by its turn, almost maximally violate the Bell's inequality. This result suggests that with a simple modification the DQC1 model can be promoted to a universal model of quantum computing.

\end{abstract}
\maketitle
\section{Introduction}
As quantum correlations are exclusive to the quantum realm, we may expect that advantages to process, store, and transmit information are somehow connected to them. In the case of processing information, to have a speedup in relation to classical algorithms, quantum computing with pure states requires entanglement \cite{jozsa2003role,van2013universal}. However, this requirement is not clear for quantum computation with mixed states. For instance, the Deterministic Quantum Computation with One Qubit (DQC1) \cite{knill1998power} is a restricted model of quantum computing using a simple setup of a single pure qubit and $n$ others in the totally mixed one. This simple scheme can evaluate the normalized trace of an arbitrary unitary matrix measuring just the first qubit. In Ref. \cite{datta2005entanglement}, Datta \emph{et al.} showed that this model requires little or null entanglement to perform the task of trace evaluation, suggesting a different quantum resource to explain the resulting computational gain \cite{datta2007role}, \emph{e.g.}, quantum discord \cite{datta2008quantum}. Notwithstanding, in Ref. \cite{dakic2010necessary}, Daki{\'c} \emph{et al.} presented a class of unitary matrices with null quantum discord without loosing the quantum advantage. Continuing this investigation, the coherence of the control qubit was pointed out as the resource responsible for the speedup, given it can be converted into other correlations \cite{ma2016converting,matera2016coherent}.

The goal of this work is to promote quantum correlations in the DQC1 model. As in the standard DQC1 model only quantum discord and coherence are present for two qubits, we use post-selection to purify the auxiliary qubit such that entanglement and Bell's nonlocality are created. 


This article is divided as follows: in Sec. \ref{sec:parax}, we introduce the DQC1 circuit and the quantum correlation measures. The post-selection process and the analysis of the results are in in Sec. \ref{sec:postselec}. Finally, our conclusions are described in Sec. \ref{sec:conc}.

\section{Quantum correlations in the standard DQC1 model}
\label{sec:parax}
\begin{figure}[h]
	\includegraphics[angle=0, width=\columnwidth]{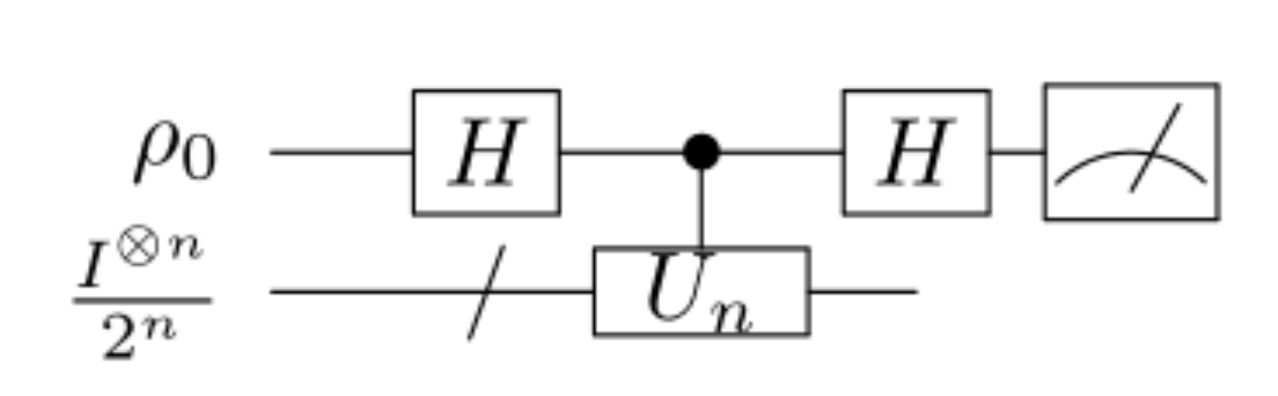}
    \caption{DQC1 circuit used to evaluate the normalized trace of a unitary matrix $U_n$. The control qubit starts in a semi-pure state $\rho_0(\alpha)=\frac{\left( I + \alpha \sigma_Z\right)}{2}$, with $0\leq \alpha \leq 1$, while the $n$ auxiliary qubits are in the maximum mixture state $\rho_n =\frac{I^{\otimes n}}{2^n}$.}
    \label{figure1}
\end{figure}

Introduced in 1998 by Knill and Laflamme \cite{knill1998power}, the Deterministic Quantum Computation with One Qubit (DQC1) is a computing model that evaluates the normalized trace of any unitary operator using a measurement in a single qubit. The DQC1 circuit (see Fig. \ref{figure1}) consists of a control qubit in the state 
\[
\rho_0(\alpha)=\frac{\left( I + \alpha \sigma_Z\right)}{2},
\]
where $\sigma_Z$ is the Pauli matrix $Z$, a certain degree of coherence controlled by $\alpha$ ($0\leq \alpha \leq 1$), and $n$ auxiliary qubits initially in the totally mixed state 
\[
\rho_n =\frac{I^{\otimes n}}{2^n},
\]
where $I$ the identity operator. 

Besides the Hadamard gates,
\[
H =\frac{1}{\sqrt{2}}
\begin{bmatrix}
1&\phantom{-}1\\
1&-1\\
\end{bmatrix},
\]
applied on the first qubit, the computation is performed by a controlled unitary between the first and auxiliary qubits. The result is obtained by measuring the expected values of the operators associated to the control qubit, whose precision is independent on the dimension of the unitary transformation $U_n$ and depends only on the number of runs of the quantum circuit \cite{datta2005entanglement}. In contrast, the simplest classical algorithm to calculate the trace of a matrix depends on its dimension, which one can increase exponentially according to number of qubits in the system \cite{datta2008studies}.  

With this advantage in mind, the idea is to map certain problems into DQC1 model. Despite of not being a universal model of quantum computing, some quantum solutions obtained through this model present advantages in comparison to their classical counterparts, such as spectral density estimation \cite{knill1998power}, testing integrability \cite{poulin2003testing}, Shor factorization \cite{parker2000efficient}, evaluation of average fidelity decay \cite{poulin2004exponential}, estimate of Jones Polynomials \cite{shor2007estimating}, and quantum metrology \cite{boixo2008parameter,cable2016power}. Inspired by these results, some experimental implementations of this model have been made in optical \cite{lanyon2008experimental, hor2015deterministic}, Nuclear Magnetic Resonance \cite{passante2009experimental,passante2011experimental}, superconducting materials \cite{wang2019witnessing}, and cold atoms \cite{krzyzanowska2017quantum}.

\subsection{Quantum Correlations in DQC1 model}

We start by writing the output state of the DQC1 circuit for two qubits in the Fano representation \cite{fano1983pairs},
\begin{multline}
\rho = \frac{1}{4} \Bigg[  I_A\otimes I_B + I_A\otimes \left(\Vec{r}\cdot\Vec{\sigma}^B\right)  \\ 
   +  \left(\Vec{s}\cdot\Vec{\sigma}^A \right)\otimes I_B + \sum_{i,j = x,y,z} c_{ij}\sigma_i^A\otimes\sigma_j^B \Bigg],   
\end{multline}
where the polarization vectors are $\Vec{r} = tr[\rho (I_A \otimes \Vec{\sigma}^B)]$ and $\Vec{s} = tr[\rho (\Vec{\sigma}^A \otimes I_B)]$, the elements of the correlation matrix $C$ are $c_{ij} = tr[\rho(\Vec{\sigma_i}^A \otimes \Vec{\sigma_j}^B)]$, the indices $A$ and $B$ refer to the first and second subsystems, and $i,j=x,y,z$ to the indices of a vector of Pauli matrices, namely $\{ \sigma_x,\sigma_y,\sigma_z\}$.

It is simple to define the quantum correlations used throughout this work (Bell's nonlocality, entanglement, quantum discord, and coherence) using this representation. 

\emph{Bell's nonlocality} - The Bell's inequality can be evaluated by the quantity \cite{horodecki1995violating}
\begin{equation}
     B(\rho) = 2\sqrt{m_1+m_2},
\end{equation}
where $m_1$ and $m_2$ are the two largest eigenvalues of the matrix $T =CC^{T}$, where $T$ means the transposition operation. If $B(\rho) \le 2$ the Bell's inequality is not violated, otherwise, non-local effects might appear. The values of $B(\rho)$ are comprised in the interval $[0,2\sqrt{2}$], with the maximum violation being achieved by the entangled pure states, such as the Bell entangled states.

\emph{Entanglement} - To quantify bipartite entanglement, we used \emph{Negativity} \cite{vidal2002computable} defined as
\begin{equation}
    N(\rho) = \frac{\|\rho^{T_A}\|_1-1}{2},
\end{equation}
where $T_A$ is the partial transposition of the subsystem $A$ and $\norm{\cdot}_1$ is the trace norm. The negativity indicates how far the partial trace of the density matrix is far from positive, and consequently, how much entangled the subsystems are. For a two-qubit system $N(\rho) \in [0,1/2]$. 

\emph{Quantum discord} - The geometric discord of an arbitrary two-qubit state is \cite{dakic2010necessary}
\begin{equation}
    D(\rho)=\frac{1}{4}\left(\|\Vec{s}\|_2^2+\|C\|_2^2-\lambda_{max}\right),
\end{equation}
where the norms in the right hand side of the equation above have been calculated using the euclidean (for vector $\Vec{s}$) and Hilbert-Schmidt (for matrix $C$) norms. $\lambda_{max}$ represents the largest eigenvalue of the matrix $$\Lambda =\Vec{s}\cdot \Vec{s}^T+CC^T.$$
The values of quantum discord are restricted to the interval $D(\rho) \in [0,1/2]$.

\emph{Coherence} - The trace norm Coherence is measured using \cite{baumgratz2014quantifying,yu2016alternative}
\begin{equation}
    C(\rho) = \norm{\rho-\rho_{diag}}_1,
\end{equation}
where $\rho_{diag}$ denotes the state obtained from $\rho$ using just the diagonal elements. Basically, this measure sums the absolute values of all off-diagonal terms of $\rho$ so that $C(\rho) \in [0,3]$ for a two-qubit system.

\begin{figure}[h]
	\includegraphics[angle=270,width=\columnwidth]{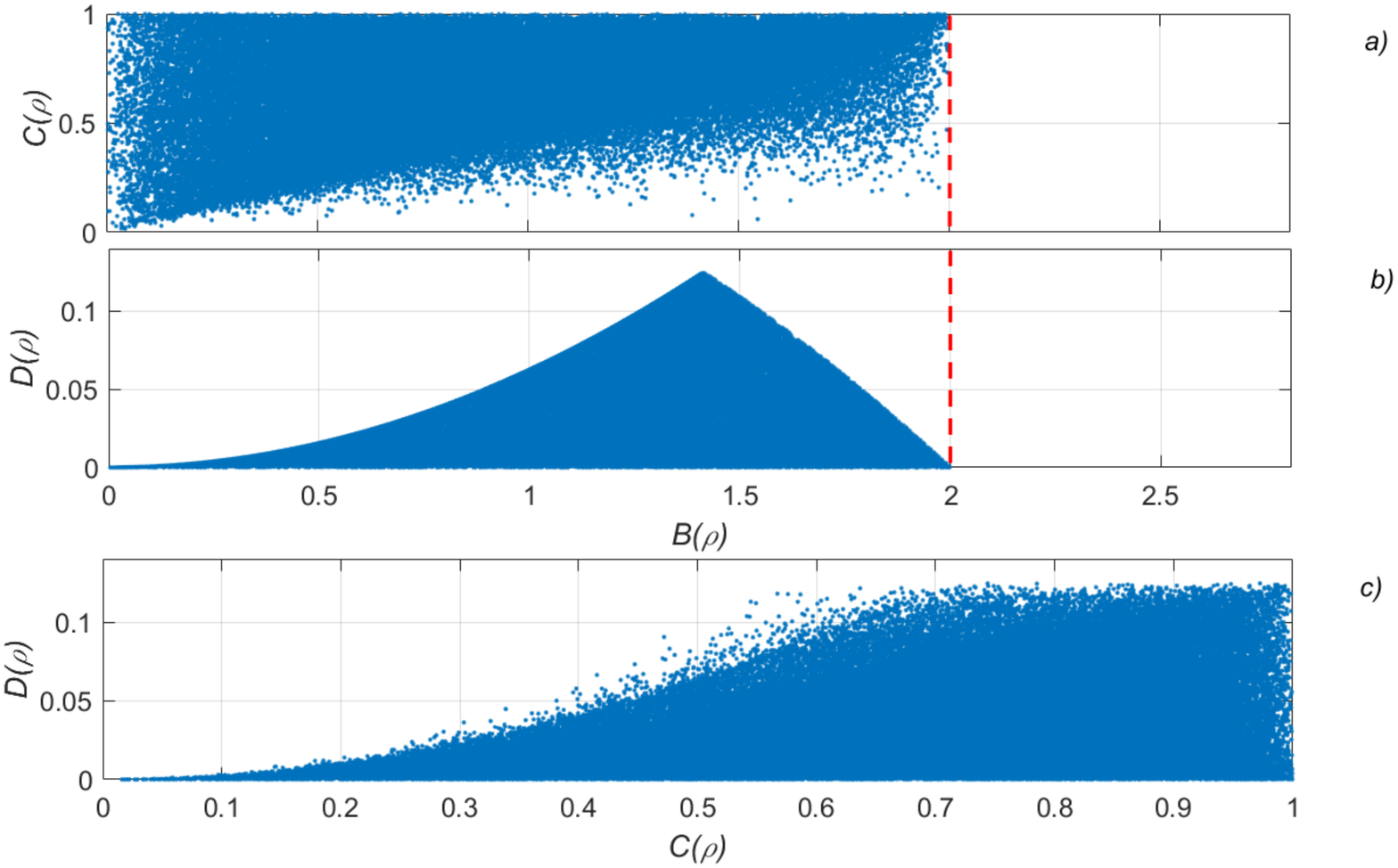}
    \caption{Each dot represents the value of the quantum correlation calculated for the output state of the DQC1 circuit. The $10^6$ states were obtained choosing initial states and unitary transformations at random. The red-dashed-vertical line represents the maximum value of $B(\rho)$ in which there is no Bell's nonlocality. The quantum correlation measured by entanglement does no appear here because it is null for the standard DQC1 circuit (see Fig. \ref{figure1}). }
    \label{figure2}
\end{figure}

To explore the quantum correlations presented in the output states of the DQC1 circuit for two qubits (see Fig. \ref{figure1}), we chose $10^6$ random initial qubit states according to Hilbert-Schmidt measure \cite{zyczkowski2011generating} and also $10^6$ unitary matrices ($U_1$) from Haar measure \cite{zyczkowski1994random}, \cite{ozols2009generate}. The quantum correlations generated between the two qubits at the end of the circuit are shown in Fig. \ref{figure2}, where each dot is obtained for a given final density matrix.

As it is already known, there is no entanglement between the control and the auxiliary qubits in the standard DQC1 model \cite{datta2005entanglement}, and a straightforward consequence of this is no violation of Bell's inequality, as shown in Figs. \ref{figure2}a and \ref{figure2}b. Superposition (quantified by coherence) is vital for the appearance of others correlations as shown in Figs. \ref{figure2}a and \ref{figure2}c. We also have, Fig. \ref{figure2}c showing states with non vanishing quantum discord $D(\rho)$ and coherence, confirming they are intimately related. Note the maximum values achieved by coherence and quantum discord, $C_{max}(\rho)=1$ and $D_{max}(\rho)=0.1244$, respectively. They are far from reaching the maximum values accessible for general two qubit states, \emph{i.e.}, $3$ and $0.5$, respectively.

\section{Post-selection in DQC1 model}
\label{sec:postselec}

Inspired by Ref. \cite{parker2000efficient}, we analyzed the effect of the post-selection on promoting quantum correlations in the DQC1 model for two qubits. 

\begin{figure}[h]
	\includegraphics[angle=0,width=\columnwidth]{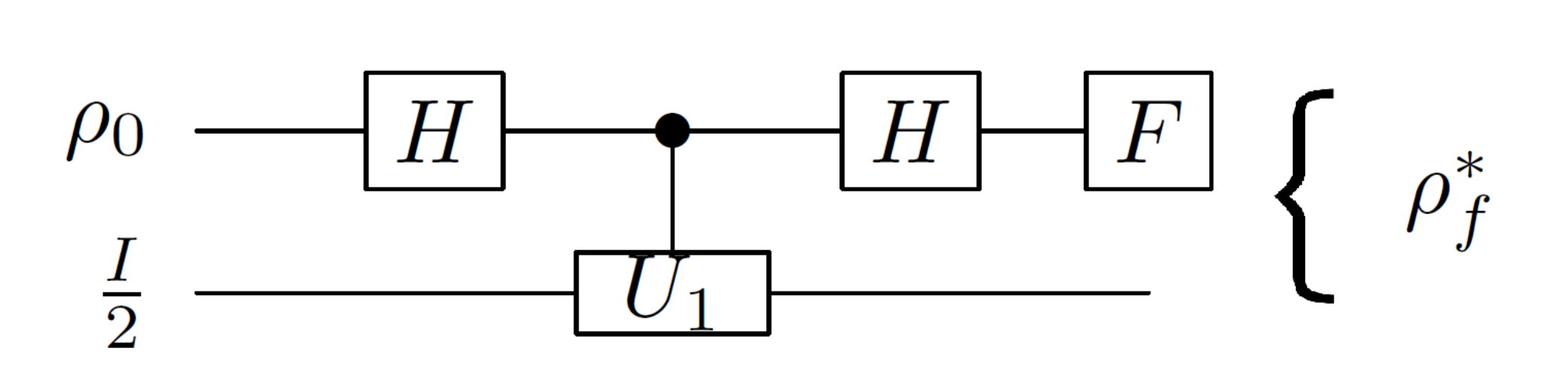}
    \caption{DQC1 model for two qubits with post-selection through a specific filter $F$ that operates on the control qubit. The initial state of the auxiliary qubit is $\rho_{n=1}=\frac{I}{2}$, while the initial state of the control qubit $\rho_0$ and the unitary $U_1$ are randomly chosen. At the end of the circuit we trace out the control qubit from the two-qubits state $\rho^*_f$ obtaining a more purified state of the auxiliary qubit. The purification process can be repeated by inserting the final state of the auxiliary qubit in the circuit again as many times as desired.}
    \label{figure3}
\end{figure}

To promote other correlations in this computing model, we introduced a specific post-selection process $F$ on the control qubit at the end of the DQC1 circuit (see Fig. \ref{figure3}), which one acts through a local filter, described by \cite{kent1999optimal}
\begin{equation}
F(U_a,\eta) = U_a
\begin{bmatrix}
1&0\\
0&\eta\\
\end{bmatrix}
U_a^{\dagger},
\end{equation}
with $\eta \in [0,1]$ and $U_a$ is a unitary matrix.  $\eta$ represents the probability of success of acting this filter considering the complete measurement $\{F(U_a,\eta), \,I-F(U_a,\eta)\}$. For the particular case $\eta = 1$, the filter reduces to the identity operator, doing nothing on the control qubit, while, for $\eta < 1$, it diminishes the contribution coming from one component of the state in the direction determined by the unitary transformation $U_a$. In the limit case $\eta=0$, just one component of the state survives.

As it is well known, if the auxiliary qubit starts in a state different from the maximum mixture, then, for an appropriate unitary $U_1$, entanglement and violation of Bell's inequality emerge in this system \cite{piani2011all}. Thus, in order to purify the auxiliary qubit, we post-select for certain values of $\eta$ and investigate the role played by this filter parameter on the quantum correlations between the qubits. 

\subsection{Benchmarking the DQC1 with post-selection}
To proceed with this investigation, we used a DQC1 with post-selection, see Fig. \ref{figure3}. The state before filtering is
\begin{equation}
    \rho_{bf} = (H \otimes I)U (H \otimes I) \rho_0 (H \otimes I) U^{\dagger} (H \otimes I),
\end{equation}
and the final state is
\begin{equation}\label{eq:rhof}
    \rho_{f}(U_a,\eta) = \frac{F(U_a,\eta)\rho_{bf}F(U_a,\eta)^\dagger}{\text{tr}\left[F(U_a,\eta)\rho_{bf}F(U_a,\eta)^\dagger\right]},
\end{equation}
for some choice of $\eta \in [0,1]$ and unitary matrix $U_a$.

Following \cite{zyczkowski2005average}, we first analyze the fidelity, 
\[
F(\rho_1,\rho_2) = \left[\sqrt{\sqrt{\rho_1}\rho_2\sqrt{\rho_1} }\right]^2,
\]
between $10^4$ pairs of output states $\rho_f$ and $\rho_f^\prime$ choosing an initial pure state $\rho_0$ for the control qubit drawn from Hilbert-Schmidt measure, a controlled unitary gate $U_1$, and a unitary $U_a$ drawn from Haar measure and fixed values of $\eta \in \{0,1/2,1\}$.

\begin{figure}[h]
	\includegraphics[angle=0,width=\columnwidth]{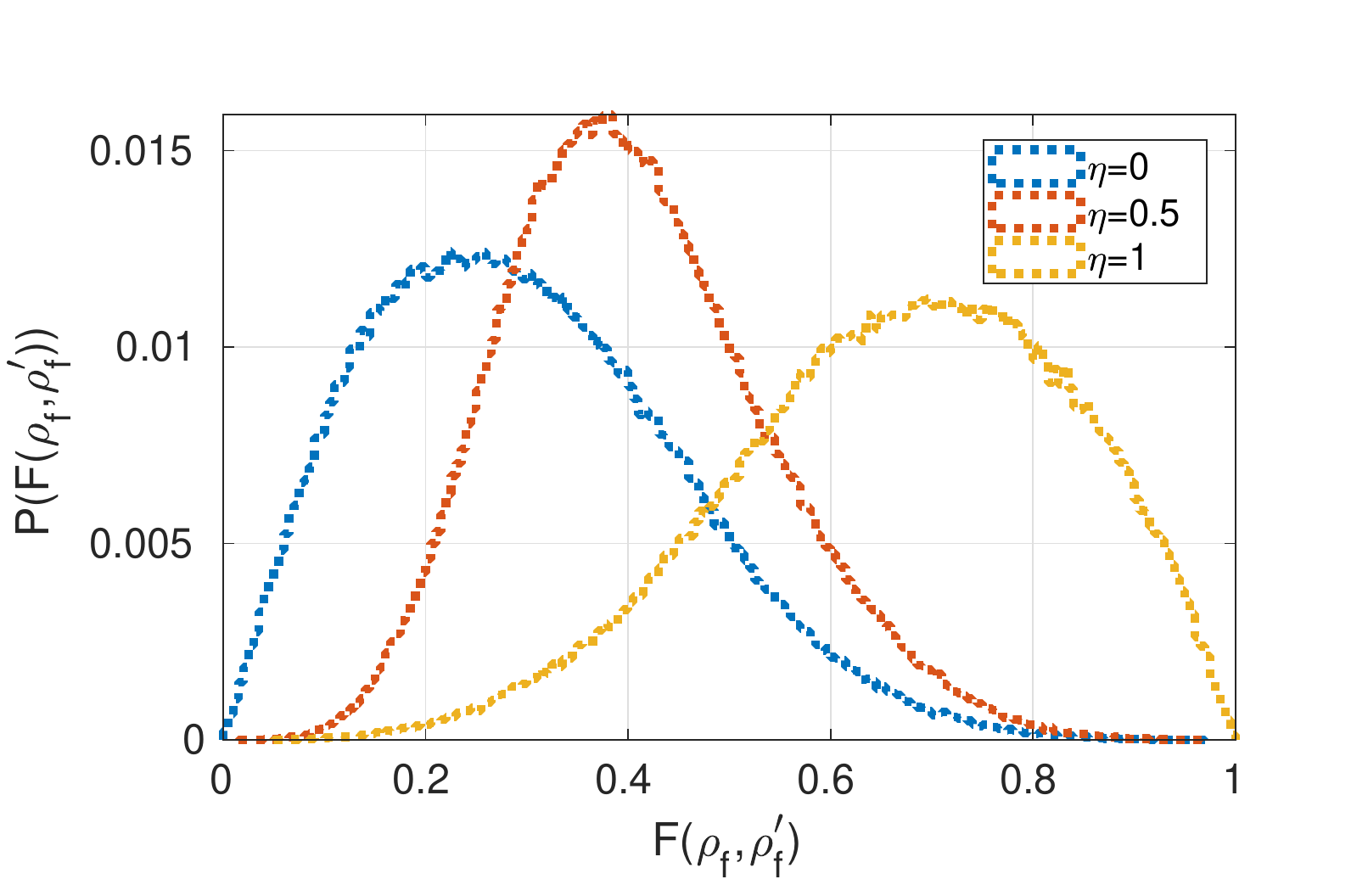}
    \caption{Emprirical probability distribution of the fidelity between $10^4$ randomly chosen $\rho_f$ and $\rho_f^\prime$ after the DQC1 with post-selection (see Eq. \ref{eq:rhof}) for fixed values of $\eta \in \{0,1/2,1\}$.}
    \label{fig:avg_fid}
\end{figure}

The probability distributions of $F(\rho_f,\rho_f^\prime)$ in Fig. \ref{fig:avg_fid} shows the average fidelity between $\rho_f$ and $\rho_f^\prime$ diminishing as $\eta \rightarrow 0$. We can see it as a numerical evidence that the DQC1 with post-selection distributes the states more distantly (according to the Bures metric \cite{wootters1981statistical}), and also with more accessible states in the two-qubit Hilbert space, which will be clear when we present the promotion of quantum correlations further in this article.

To understand qualitatively this result let us remember that the motivation of the DQC1 model relies on Nuclear Magnetic Resonance systems, whose two-qubit density matrices have the form
\begin{equation}
    \rho_{NMR} = \frac{1-\epsilon}{4}I_{4 \times 4}+\epsilon |\psi \rangle \langle \psi |,
\end{equation}
with $0 \le \epsilon \le 1$ and $I_{4 \times 4}$ and $|\psi \rangle$ are the identity operator and a pure state in the two-qubit Hilbert space. If we calculate the fidelity for states of the form $\rho_{NMR}$ it is easy to verify that for small values of $\epsilon$ states with fidelities closer to one are more frequent, which is similar to the standard DQC1 model.

\subsection{Purity and quantum correlations in DQC1 model}

\begin{figure}[h]
	\includegraphics[angle=0,width=\columnwidth]{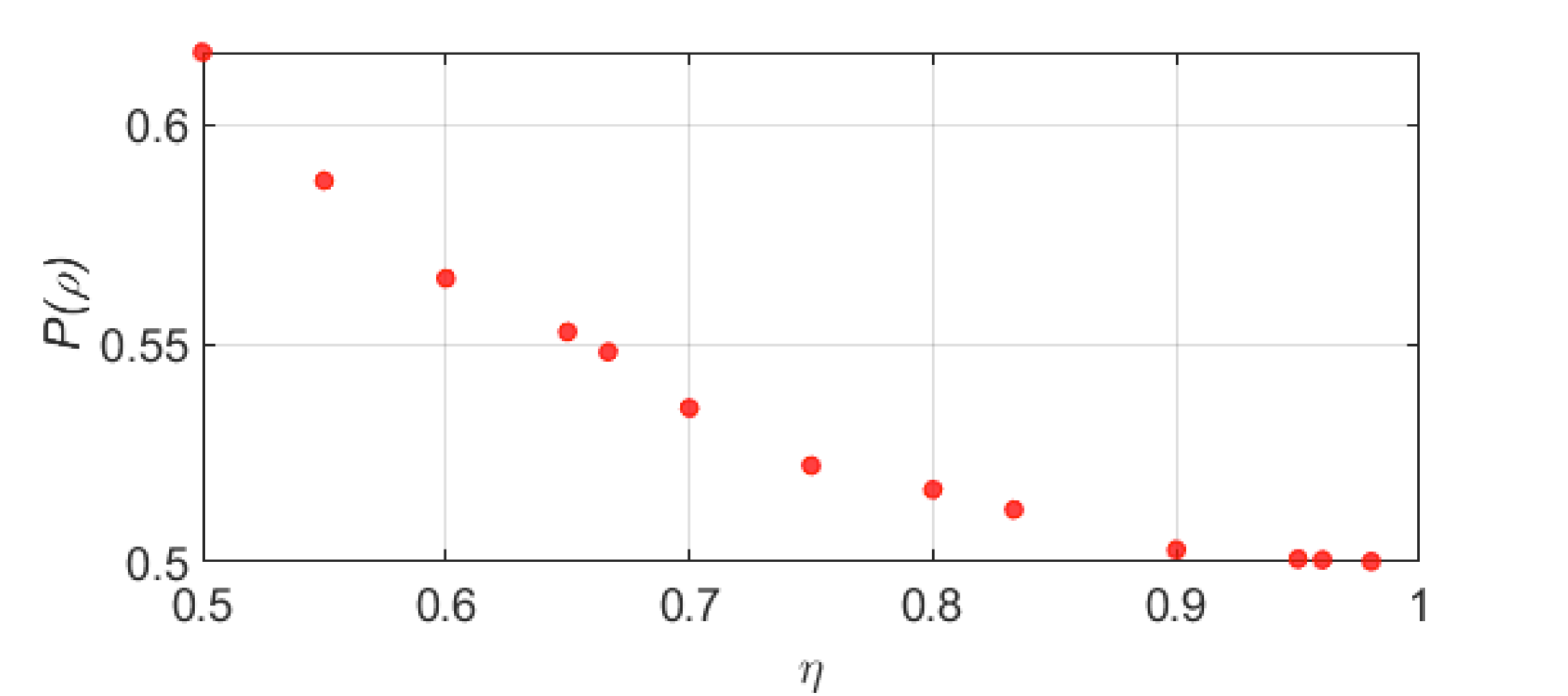}
    \caption{Purity as function of the probability of success $\eta$ for DQC1 model with two qubits and post-selection. Each dot represents the average maximum value of the achieved purity for the auxiliary qubit with a specific $\eta$.}
    \label{figure10}
\end{figure}

As mentioned before, the purity $P(\rho) = \tr \rho^2$ of the auxiliary qubit also determines the possibility of promoting quantum correlations in DQC1 model. The purity varies from $P(\rho)= 1/2$ for a totally mixed qubit state to $P(\rho)=1$ for a pure qubit state. In Fig. \ref{figure10} we run the same protocol above and take the average maximum value of the purity for a specific value of $\eta$.  We observe that the average maximum value of the purity is approximately $P(\rho) \simeq 0.62$ for $\eta = 0.5$ and as $\eta$ approaches $1$ the purity converges to its minimum value, as expected, since for $\eta = 1$ the filter has no effect on the auxiliary qubit.  

\begin{figure}[h]
	\includegraphics[angle=0,width=\columnwidth]{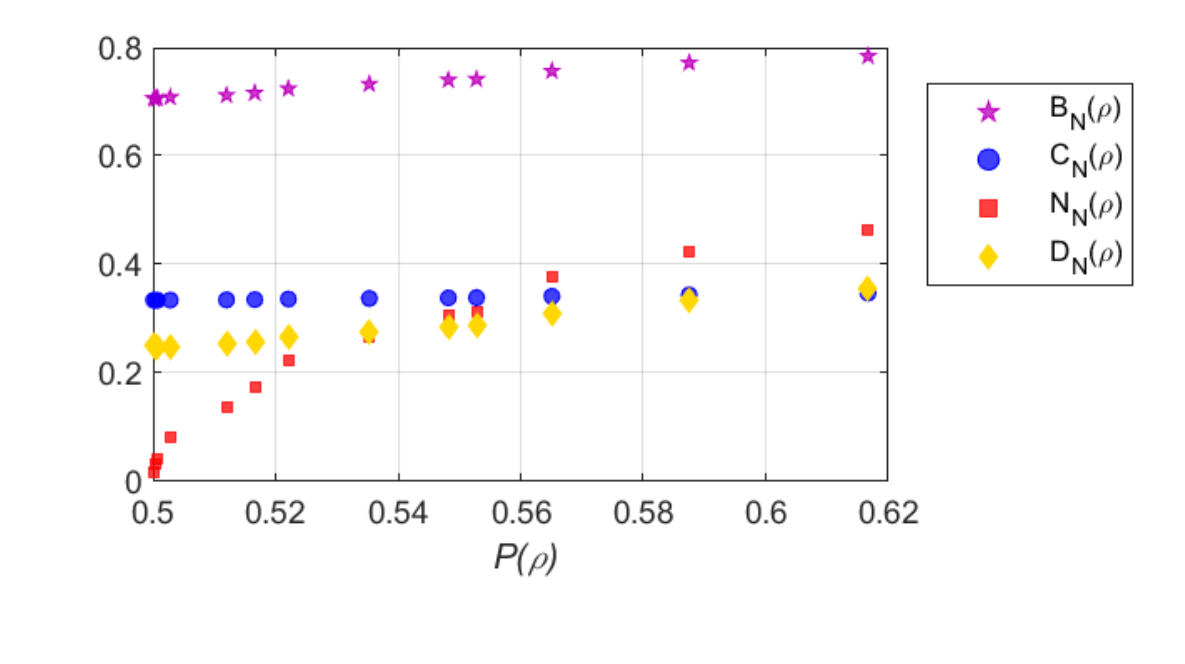}
    \caption{Normalized quantum correlations as function of purity $P(\rho)$. Each dot represents the average maximum value of the quantum correlations for a given purity.}
    \label{figure11}
\end{figure}

Figure \ref{figure11} shows the behavior of each normalized quantum correlation as function of the purity of the auxiliary qubit. The normalization of a given quantum correlation $X$ is defined as $X_N \equiv X/X_{max}$, where $X_{max}$ is the maximum value achieved by the quantum correlation $X$. All correlations increase as the purity increase. We highlight that even without a significant purification of the auxiliary qubit we already achieve non-null entanglement and Bell's nonlocality ($B_N(\rho) > 1/\sqrt{2} \simeq 0.707$).  These results are in agreement with the discussion made in the previous section in which coherence (blue discs) and quantum discord (yellow diamonds) exists independently on the purification process, once they are created by local operations (and classical communications for discord). On the other hand, the existence of entanglement (red squares) and Bell's nonlocality (pink stars) depends strongly on the purity of the auxiliary qubit. Entanglement starts from zero for totally mixed states ($P(\rho)=1/2$) and increases, but it does not achieve the maximum entanglement. A similar behavior is found for the violation of Bell's inequality, although its minimum value is greater than $1/\sqrt{2}$.

\begin{figure}[h]
	\centering
	\includegraphics[angle=0,width=\columnwidth]{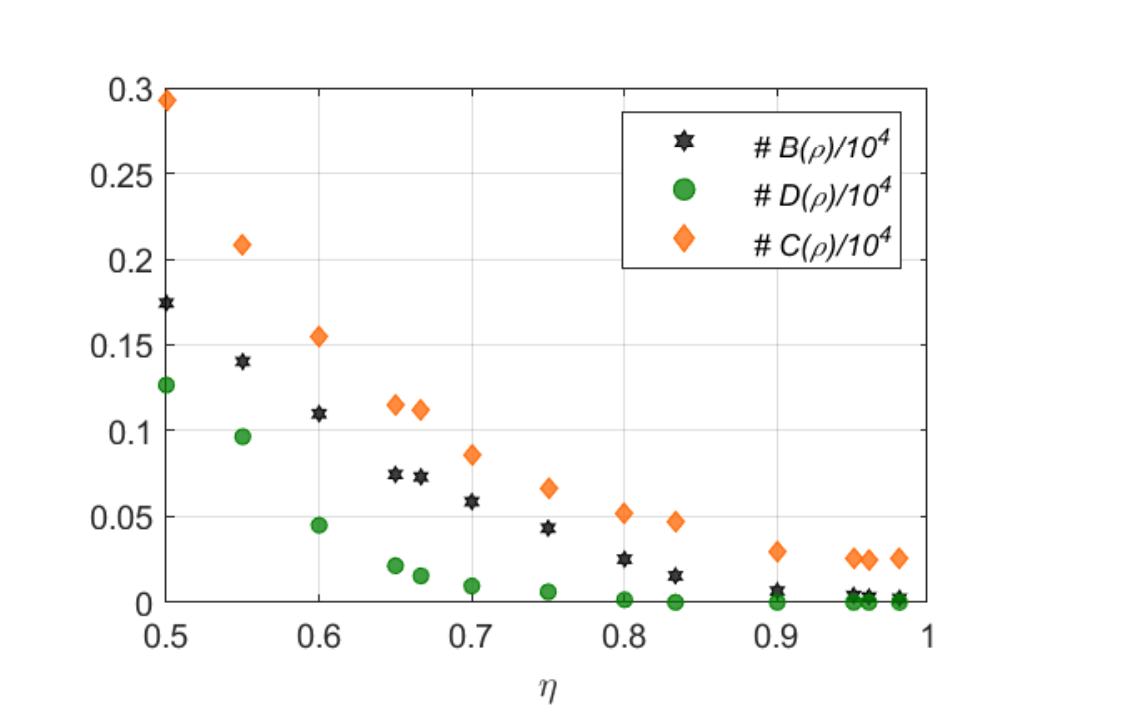}
    \caption{Density of states, defined as the ratio between the number of states after the post-selection process with a correlation  value  greater  than  the  maximum  value  of  the correlation achieved by the states in the standard DQC1model  (without  post-selection), versus the probability of success $\eta$ of the purification protocol. The total number of states analyzed for each correlation is $10^4$.}
    \label{figure18}
\end{figure}

To estimate the number of states that have been promoted after the post-selection process, \emph{i.e.}, the number of states whose quantum correlations increased after the filtering procedure, we analyse the density of states. The density of states is defined as the ratio between the number of states after the post-selection process with a correlation value greater than the maximum value of the correlation achieved by the states in the standard DQC1 model (without post-selection). The maximum values of quantum discord, quantum coherence, and $B(\rho)$ attained by the standard DQC1 circuit (see Fig. \ref{figure2}) are $0.1244$, $0.9992$, and $1.9974$, respectively. Figure \ref{figure18} presents density of states for quantum discord (green discs), quantum coherence (orange diamonds), and $B(\rho)$ (black stars) for the different values of $\eta$ considering a total of $10^4$ states. As there is no entanglement between the control and the auxiliary qubit in the standard DQC1 model, the density of states defined above does not apply for this correlation. We observe that the higher the $\eta$ the lower the density of states that overcome the value of these correlations for DQC1 model without post-selection.  This reinforce the strong connection between the mixedness of the auxiliary qubits and the quantum correlations in the model. As in the standard DQC1 model the maximum value achieved by $B(\rho)$ is approximately $2$, all states used to build Fig. \ref{figure18} (black stars) violate the Bell's inequality. These plots give us information about the number of states that are accessible after the post-selection process, which ones constitute a resource for quantum computation.

\subsection{Purification optimization}

Now we want to find optimal strategies of filtering, without fixing the parameter $\eta$, to reach a purity of $P(\rho)=0.99$ for the auxiliary system. 

For this optimization, we chose the following procedure:
\begin{enumerate}
    \item The initial pure state $\rho_0$ for the control qubit and the controlled unitary gate $U_1$ are drawn from Hilbert-Schmidt and Haar measures, respectively.
    
    \item The parameters of the filter $F$, which include $\eta$ and the unitary $U_a$, are chosen such that 
    \[ 
    \text{arg}\min_{\eta,U_a}\{ P( \text{tr}_{c} \left[ \rho_{f}(U_a,\eta) \right])\},
    \]
    where $\text{tr}_{c} \left[ \rho_{f}(U_a,\eta) \right]$ is the state of the auxiliary system after the post-selection, tracing the control system out.
   
   \item If  $P( \text{tr}_{c} \left[ \rho_{f}(U_a,\eta) \right]) \approx 0.99$ we stop.
   
    \item Otherwise, the purified state of the auxiliary qubit is chosen in the DQC1 circuit instead of the identity and we optimize again for the same choice of initial pure state $\rho_0$ for the control qubit and the controlled unitary gate $U_1$.
\end{enumerate}

\begin{figure}[h]
	\centering
	\includegraphics[angle=0,width=\columnwidth]{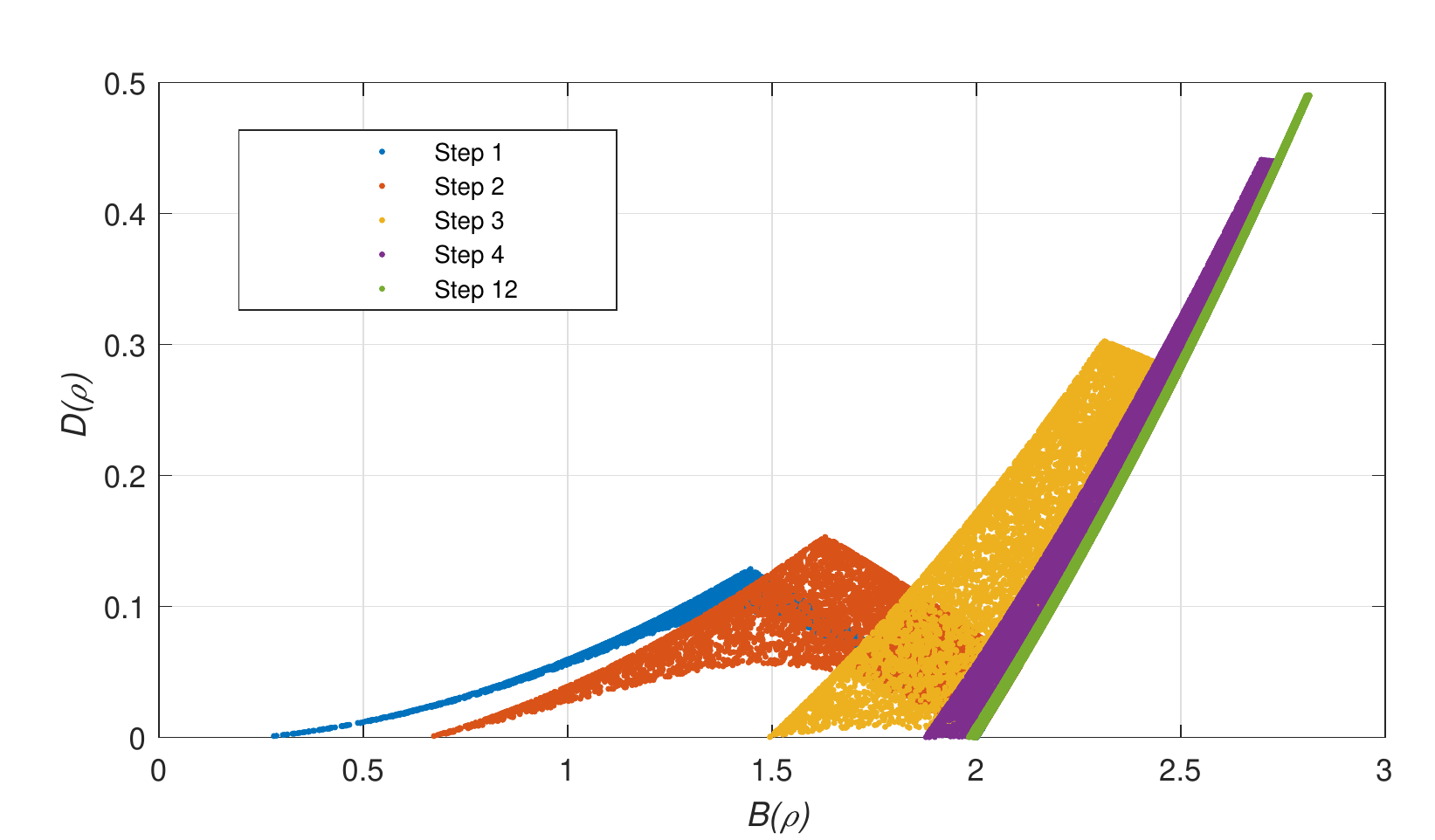}
    \caption{Quantum discord $D(\rho)$ \emph{versus} $B(\rho)$. Each surface with different color shows the mean value of $10^4$ random unitary matrices $U_1$ and control qubit initial states used to compute the quantum correlations for each purification step of the auxiliary qubit. Due to the similar behavior of the quantum correlations for the intermediate steps of the purification process only five of a total of twelve steps have been plotted.}
    \label{figure12}
\end{figure}

\begin{figure}[h]
	\centering
	\includegraphics[angle=0,width=\columnwidth]{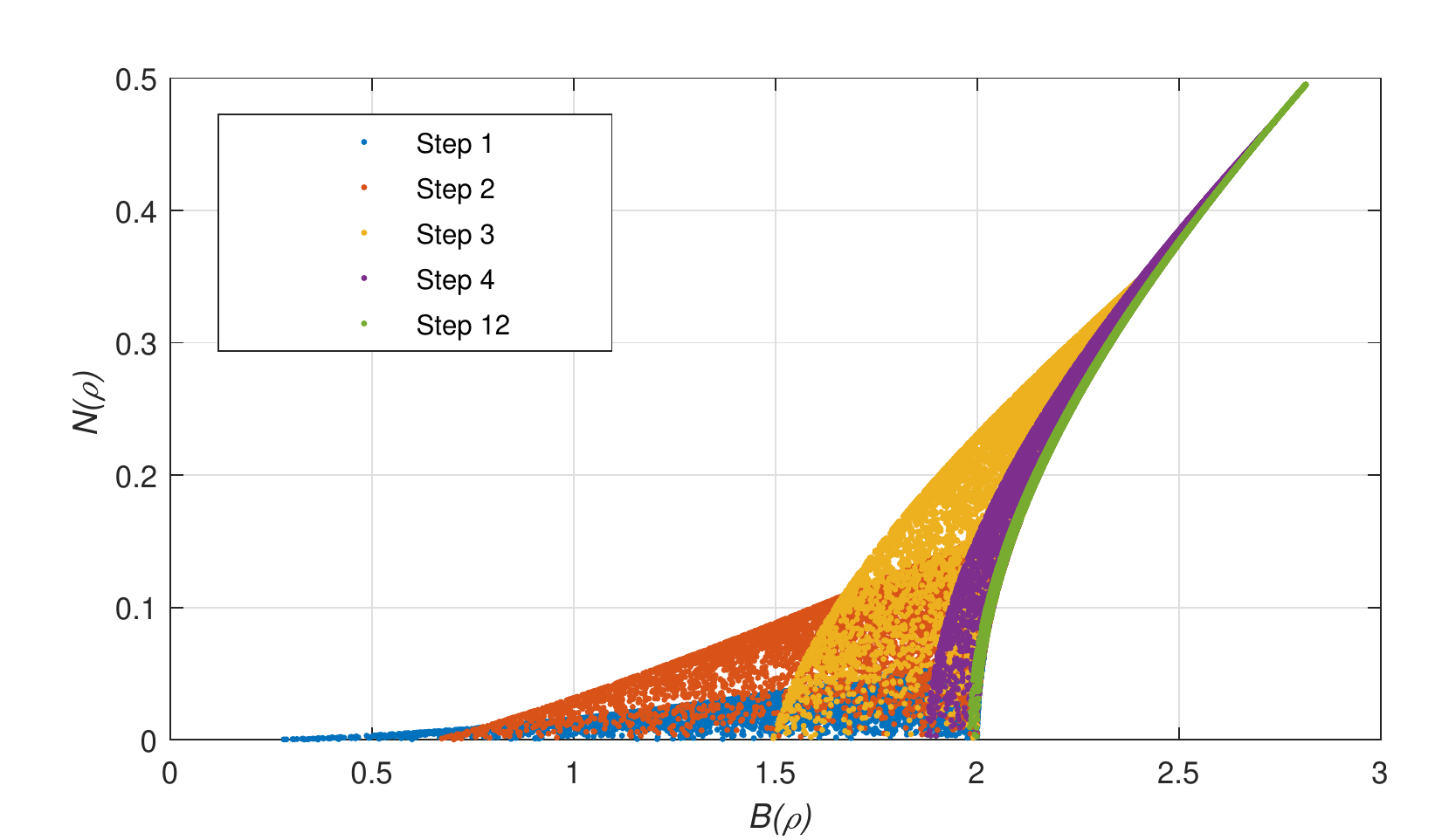}
    \caption{Negativity $N(\rho)$ \emph{versus} $B(\rho)$. Each surface with different color shows the mean value of $10^4$ random unitary matrices $U_1$ and control qubit initial states used to compute the quantum correlations for each purification step of the auxiliary qubit. Due to the similar behavior of the quantum correlations for the intermediate steps of the purification process only five of a total of twelve steps have been plotted.}
    \label{figure13}
\end{figure}

\begin{figure}[h]
	\centering
	\includegraphics[angle=0,width=\columnwidth]{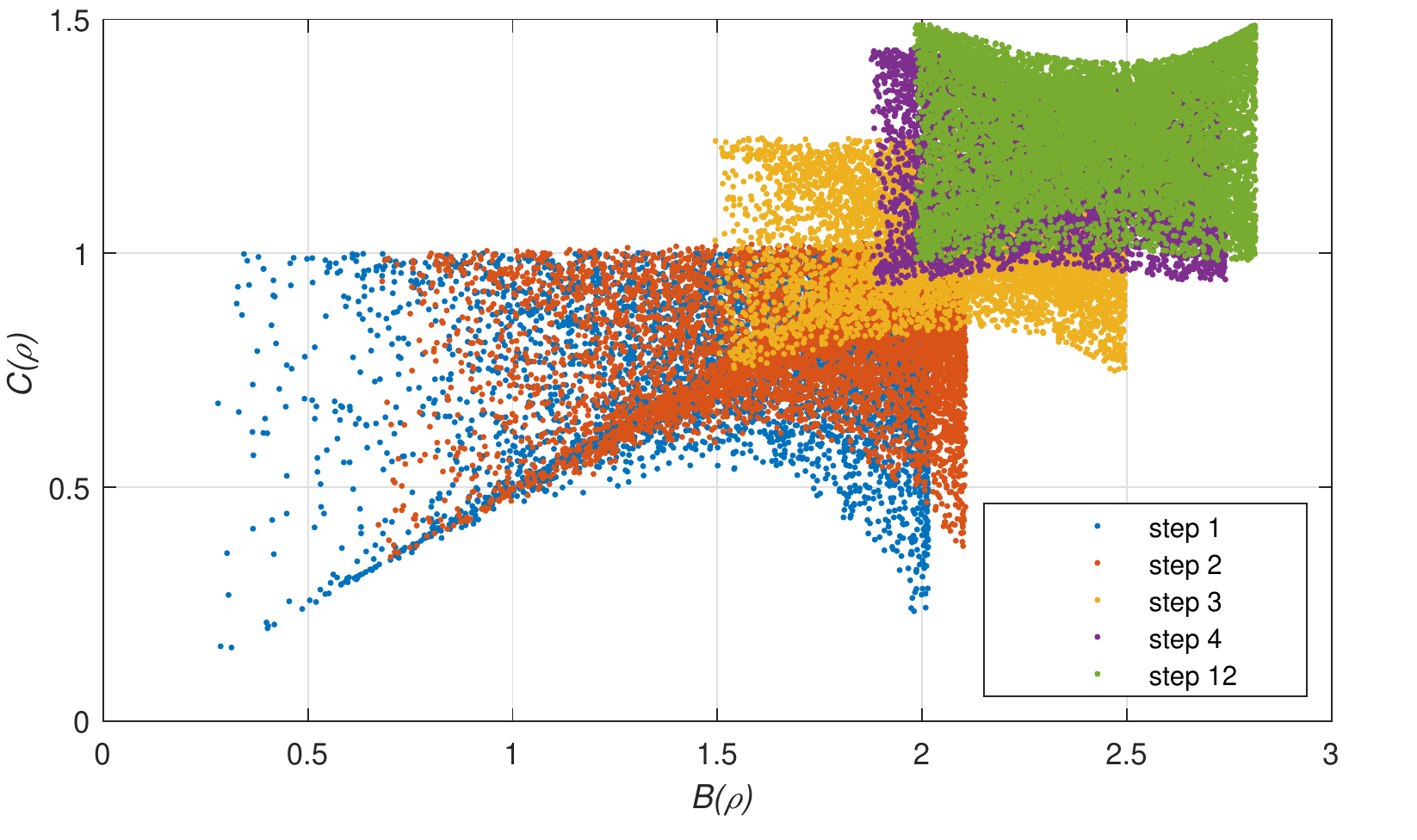}
    \caption{Quantum coherence $C(\rho)$ \emph{versus} $B(\rho)$. Each surface with different color shows the mean value of $10^4$ random unitary matrices $U_1$ and control qubit initial states used to compute the quantum correlations for each purification step of the auxiliary qubit. Due to the similar behavior of the quantum correlations for the intermediate steps of the purification process only five of a total of twelve steps have been plotted.}
    \label{figure14}
\end{figure}

\begin{figure}[h]
	\includegraphics[angle=0,width=\columnwidth]{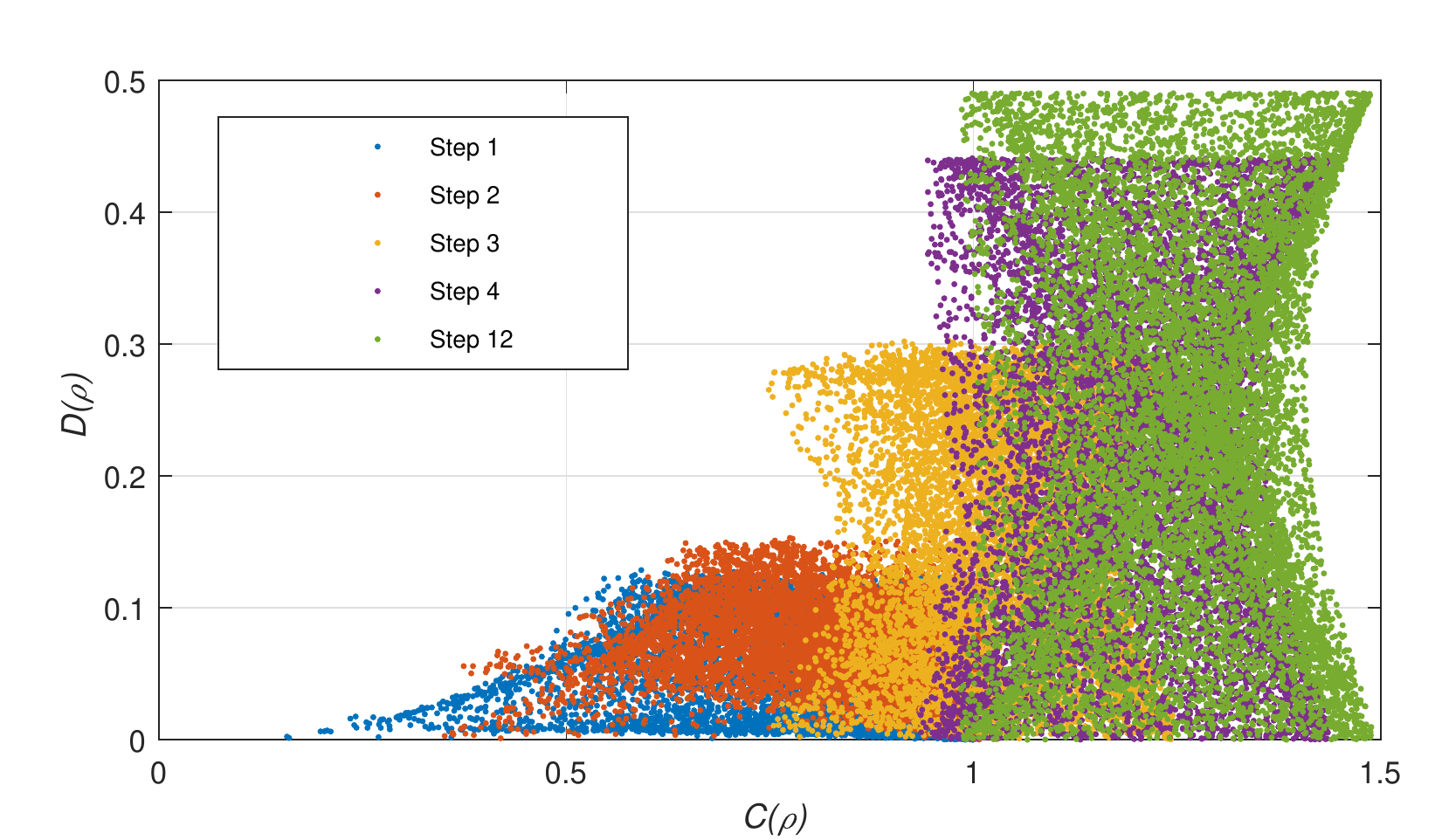}
    \caption{Quantum discord $D(\rho)$ \emph{versus} quantum coherence $C(\rho)$. Each surface with different color shows the mean value of $10^4$ random unitary matrices $U_1$ and control qubit initial states used to compute the quantum correlations for each purification step of the auxiliary qubit. Due to the similar behavior of the quantum correlations for the intermediate steps of the purification process only five of a total of twelve steps have been plotted.}
    \label{figure15}
\end{figure}

\begin{figure}[h]
	\centering
	\includegraphics[angle=0,width=\columnwidth]{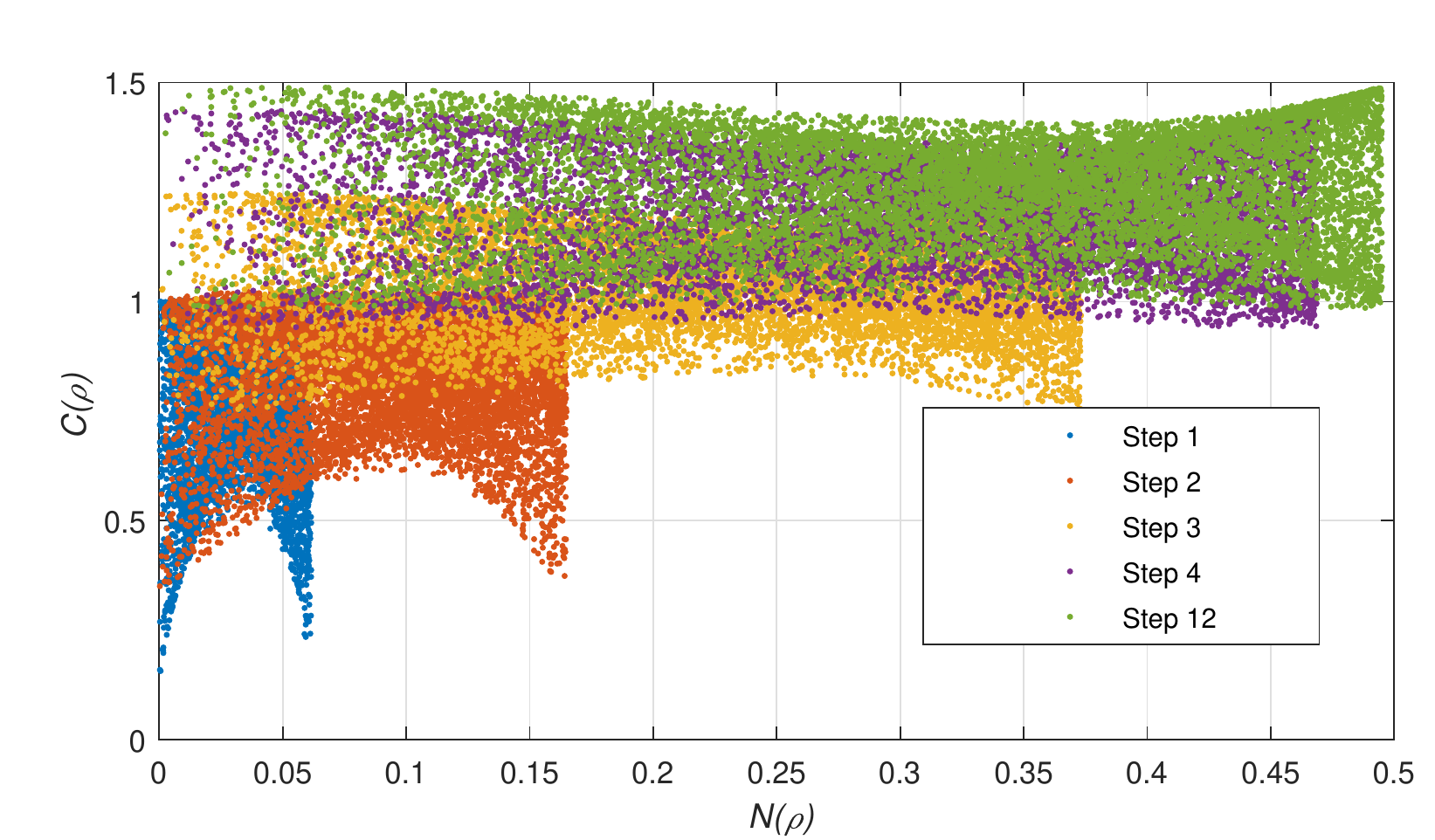}
    \caption{Quantum coherence $C(\rho)$ \emph{versus} negativity $N(\rho)$. Each surface with different color shows the mean value of $10^4$ random unitary matrices $U_1$ and control qubit initial states used to compute the quantum correlations for each purification step of the auxiliary qubit. Due to the similar behavior of the quantum correlations for the intermediate steps of the purification process only five of a total of twelve steps have been plotted.}
    \label{figure16}
\end{figure}

\begin{figure}[h]
	\centering
	\includegraphics[angle=0,width=\columnwidth]{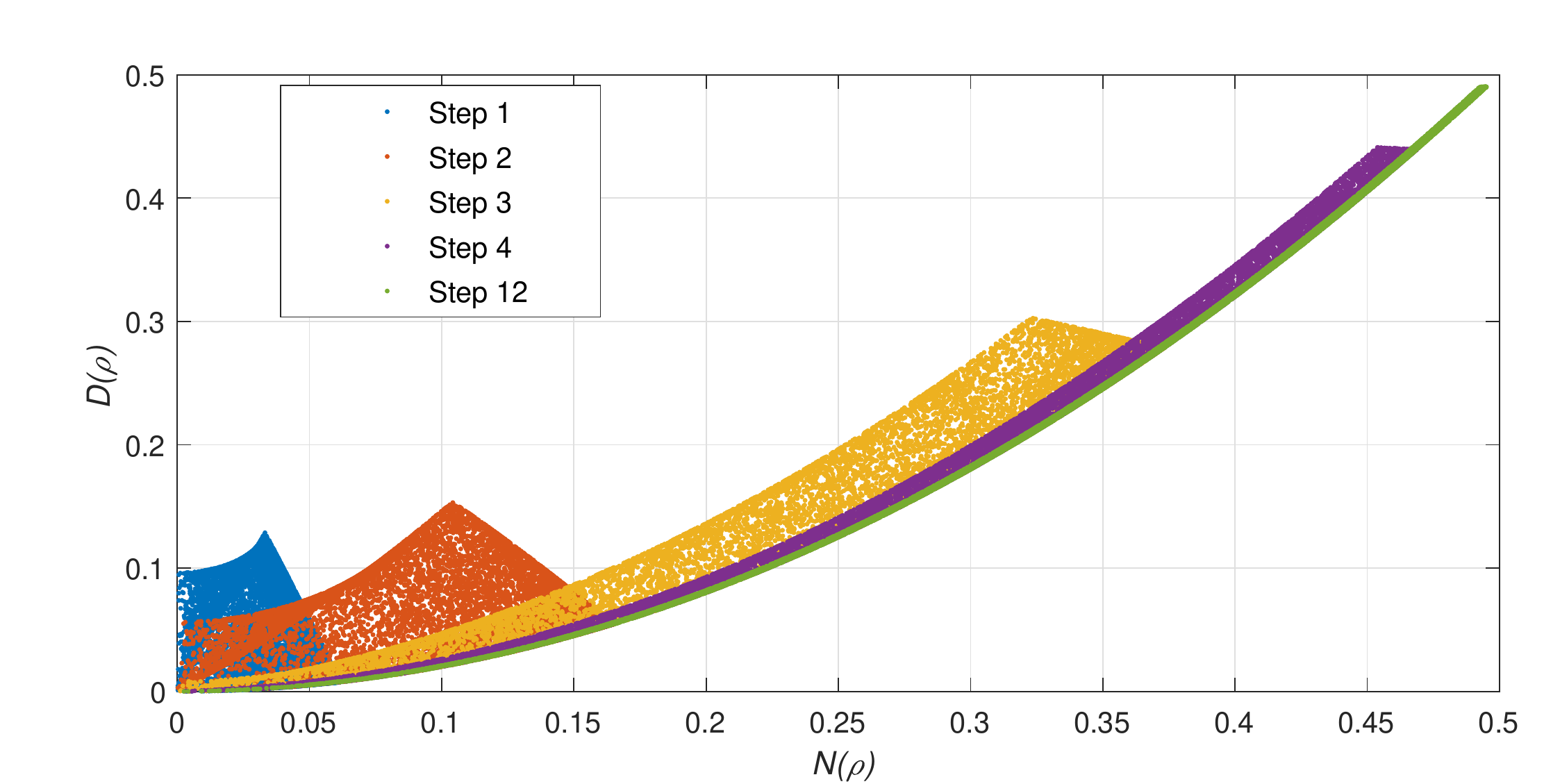}
    \caption{Quantum discord $D(\rho)$ \emph{versus} negativity $N(\rho)$. Each surface with different color shows the mean value of $10^4$ random unitary matrices $U_1$ and control qubit initial states used to compute the quantum correlations for each purification step of the auxiliary qubit. Due to the similar behavior of the quantum correlations for the intermediate steps of the purification process only five of a total of twelve steps have been plotted.}
    \label{figure17}
\end{figure}

We noticed it was necessary $12$ steps on average to achieve the desired value of purity. The relation between the quantum correlations with the steps of purification are shown in Figures \ref{figure12}-\ref{figure17}. Each surface with different color shows the mean value of $10^4$ random unitary matrices $U_1$ and initial states of the control qubit used to compute the quantum correlations for each purification step of the auxiliary qubit. In order to keep the figures legible, we have plotted only $5$ steps of a total of $12$, \emph{i.e.}, the first, second, third, fourth, and twelfth ones. The blue surfaces represent the mean value of quantum correlations after the first step of purification, while the green ones after the last step of purification, in which the auxiliary qubit achieves $P(\rho)=0.99$. 

Figures \ref{figure12}, \ref{figure13}, and \ref{figure14} show that according to the increase of the purity of the auxiliary qubit, the number of states that violates the Bell's inequality also increases almost achieving its maximum violation. From step three onward the quantum correlations behave totally different. In Fig. \ref{figure14} we clearly see that the Bell's inequality is violated already in the second step of purification and from the fourth step onward almost all states are nonlocal. Also, Figs. \ref{figure12} and \ref{figure13} agree for highly purified auxiliary states once quantum discord and entanglement quantify the same kind of correlation \cite{dakic2010necessary}.

In Fig. \ref{figure14}, we observe that in all steps of purification of the auxiliary qubit, the states have non-null coherence, achieving the maximum value of approximately $1.5$. Such a value is also achieved by states that violate almost maximally the Bell's inequality, \emph{i.e.}, they are of the form $\frac{|00\rangle + |11\rangle}{\sqrt{2}}$, which gives us a coherence value of $\sqrt{2} \simeq 1.41$.

Figures \ref{figure15} and \ref{figure16} show that for the last step of purification, almost all states have $C(\rho)>1$, which demonstrate the efficiency of the purification protocol.

The general behavior of quantum correlations presented in Fig. \ref{figure17} agrees with the result presented in Ref. \cite{debarba2012witnessed}, which one states that when using $2$-norm, quantum discord is lower bounded by negativity. As mentioned above, for entangled states almost pure, both measure become quite similar.

\section{Conclusions}
\label{sec:conc}

In this work we showed how to promote quantum correlations, such as coherence, quantum discord, entanglement, and Bell's nonlocality, present in the output state of the DQC1 model for two qubits. As already known, in the standard DQC1 model for two qubits there are only quantum discord and coherence. By applying a filtering process combined to optimization protocols, entanglement and Bell's nonlocality arise in this system, even for a small purification of the auxiliary qubit. In the case we reintroduced the purified auxiliary qubit into the circuit again, we observed that the number of purification steps that are needed to achieve a maximum purity of $0.99$ is on average $12$ steps. For this level of purity the qubits become practically maximally entangled and violate the Bell's inequality almost maximally too. Our results suggest that it is possible to promote this restricted model of quantum computing to a universal one by using post-selection with a specific filter, although a formal proof is lacking.

\label{sec:conclusions}

\begin{acknowledgments}
The authors thank Ana Sprotte, Ernesto Galvão, and Susane Caligari for useful discussions. The authors also acknowledge the financial support from the Brazilian funding agencies CNPq, CAPES, and the National Institute for Science and Technology of Quantum Information (INCT/IQ).

\end{acknowledgments}

\section{References}
\bibliography{bibliografia}

\begin{thebibliography}{35}
\expandafter\ifx\csname natexlab\endcsname\relax\def\natexlab#1{#1}\fi
\expandafter\ifx\csname bibnamefont\endcsname\relax
  \def\bibnamefont#1{#1}\fi
\expandafter\ifx\csname bibfnamefont\endcsname\relax
  \def\bibfnamefont#1{#1}\fi
\expandafter\ifx\csname citenamefont\endcsname\relax
  \def\citenamefont#1{#1}\fi
\expandafter\ifx\csname url\endcsname\relax
  \def\url#1{\texttt{#1}}\fi
\expandafter\ifx\csname urlprefix\endcsname\relax\def\urlprefix{URL }\fi
\providecommand{\bibinfo}[2]{#2}
\providecommand{\eprint}[2][]{\url{#2}}

\bibitem[{\citenamefont{Jozsa and Linden}(2003)}]{jozsa2003role}
\bibinfo{author}{\bibfnamefont{R.}~\bibnamefont{Jozsa}} \bibnamefont{and}
  \bibinfo{author}{\bibfnamefont{N.}~\bibnamefont{Linden}},
  \bibinfo{journal}{Proceedings of the Royal Society of London. Series A:
  Mathematical, Physical and Engineering Sciences}
  \textbf{\bibinfo{volume}{459}}, \bibinfo{pages}{2011} (\bibinfo{year}{2003}).

\bibitem[{\citenamefont{Van~den Nest}(2013)}]{van2013universal}
\bibinfo{author}{\bibfnamefont{M.}~\bibnamefont{Van~den Nest}},
  \bibinfo{journal}{Physical review letters} \textbf{\bibinfo{volume}{110}},
  \bibinfo{pages}{060504} (\bibinfo{year}{2013}).

\bibitem[{\citenamefont{Knill and Laflamme}(1998)}]{knill1998power}
\bibinfo{author}{\bibfnamefont{E.}~\bibnamefont{Knill}} \bibnamefont{and}
  \bibinfo{author}{\bibfnamefont{R.}~\bibnamefont{Laflamme}},
  \bibinfo{journal}{Physical Review Letters} \textbf{\bibinfo{volume}{81}},
  \bibinfo{pages}{5672} (\bibinfo{year}{1998}).

\bibitem[{\citenamefont{Datta et~al.}(2005)\citenamefont{Datta, Flammia, and
  Caves}}]{datta2005entanglement}
\bibinfo{author}{\bibfnamefont{A.}~\bibnamefont{Datta}},
  \bibinfo{author}{\bibfnamefont{S.~T.} \bibnamefont{Flammia}},
  \bibnamefont{and} \bibinfo{author}{\bibfnamefont{C.~M.} \bibnamefont{Caves}},
  \bibinfo{journal}{Physical Review A} \textbf{\bibinfo{volume}{72}},
  \bibinfo{pages}{042316} (\bibinfo{year}{2005}).

\bibitem[{\citenamefont{Datta and Vidal}(2007)}]{datta2007role}
\bibinfo{author}{\bibfnamefont{A.}~\bibnamefont{Datta}} \bibnamefont{and}
  \bibinfo{author}{\bibfnamefont{G.}~\bibnamefont{Vidal}},
  \bibinfo{journal}{Physical Review A} \textbf{\bibinfo{volume}{75}},
  \bibinfo{pages}{042310} (\bibinfo{year}{2007}).

\bibitem[{\citenamefont{Datta et~al.}(2008)\citenamefont{Datta, Shaji, and
  Caves}}]{datta2008quantum}
\bibinfo{author}{\bibfnamefont{A.}~\bibnamefont{Datta}},
  \bibinfo{author}{\bibfnamefont{A.}~\bibnamefont{Shaji}}, \bibnamefont{and}
  \bibinfo{author}{\bibfnamefont{C.~M.} \bibnamefont{Caves}},
  \bibinfo{journal}{Physical review letters} \textbf{\bibinfo{volume}{100}},
  \bibinfo{pages}{050502} (\bibinfo{year}{2008}).

\bibitem[{\citenamefont{Daki{\'c} et~al.}(2010)\citenamefont{Daki{\'c}, Vedral,
  and Brukner}}]{dakic2010necessary}
\bibinfo{author}{\bibfnamefont{B.}~\bibnamefont{Daki{\'c}}},
  \bibinfo{author}{\bibfnamefont{V.}~\bibnamefont{Vedral}}, \bibnamefont{and}
  \bibinfo{author}{\bibfnamefont{{\v{C}}.}~\bibnamefont{Brukner}},
  \bibinfo{journal}{Physical review letters} \textbf{\bibinfo{volume}{105}},
  \bibinfo{pages}{190502} (\bibinfo{year}{2010}).

\bibitem[{\citenamefont{Ma et~al.}(2016)\citenamefont{Ma, Yadin, Girolami,
  Vedral, and Gu}}]{ma2016converting}
\bibinfo{author}{\bibfnamefont{J.}~\bibnamefont{Ma}},
  \bibinfo{author}{\bibfnamefont{B.}~\bibnamefont{Yadin}},
  \bibinfo{author}{\bibfnamefont{D.}~\bibnamefont{Girolami}},
  \bibinfo{author}{\bibfnamefont{V.}~\bibnamefont{Vedral}}, \bibnamefont{and}
  \bibinfo{author}{\bibfnamefont{M.}~\bibnamefont{Gu}},
  \bibinfo{journal}{Physical review letters} \textbf{\bibinfo{volume}{116}},
  \bibinfo{pages}{160407} (\bibinfo{year}{2016}).

\bibitem[{\citenamefont{Matera et~al.}(2016)\citenamefont{Matera, Egloff,
  Killoran, and Plenio}}]{matera2016coherent}
\bibinfo{author}{\bibfnamefont{J.~M.} \bibnamefont{Matera}},
  \bibinfo{author}{\bibfnamefont{D.}~\bibnamefont{Egloff}},
  \bibinfo{author}{\bibfnamefont{N.}~\bibnamefont{Killoran}}, \bibnamefont{and}
  \bibinfo{author}{\bibfnamefont{M.~B.} \bibnamefont{Plenio}},
  \bibinfo{journal}{Quantum Science and Technology}
  \textbf{\bibinfo{volume}{1}}, \bibinfo{pages}{01LT01} (\bibinfo{year}{2016}).

\bibitem[{\citenamefont{Datta}(2008)}]{datta2008studies}
\bibinfo{author}{\bibfnamefont{A.}~\bibnamefont{Datta}},
  \bibinfo{journal}{arXiv preprint arXiv:0807.4490}  (\bibinfo{year}{2008}).

\bibitem[{\citenamefont{Poulin et~al.}(2003)\citenamefont{Poulin, Laflamme,
  Milburn, and Paz}}]{poulin2003testing}
\bibinfo{author}{\bibfnamefont{D.}~\bibnamefont{Poulin}},
  \bibinfo{author}{\bibfnamefont{R.}~\bibnamefont{Laflamme}},
  \bibinfo{author}{\bibfnamefont{G.}~\bibnamefont{Milburn}}, \bibnamefont{and}
  \bibinfo{author}{\bibfnamefont{J.~P.} \bibnamefont{Paz}},
  \bibinfo{journal}{Physical Review A} \textbf{\bibinfo{volume}{68}},
  \bibinfo{pages}{022302} (\bibinfo{year}{2003}).

\bibitem[{\citenamefont{Parker and Plenio}(2000)}]{parker2000efficient}
\bibinfo{author}{\bibfnamefont{S.}~\bibnamefont{Parker}} \bibnamefont{and}
  \bibinfo{author}{\bibfnamefont{M.~B.} \bibnamefont{Plenio}},
  \bibinfo{journal}{Physical review letters} \textbf{\bibinfo{volume}{85}},
  \bibinfo{pages}{3049} (\bibinfo{year}{2000}).

\bibitem[{\citenamefont{Poulin et~al.}(2004)\citenamefont{Poulin, Blume-Kohout,
  Laflamme, and Ollivier}}]{poulin2004exponential}
\bibinfo{author}{\bibfnamefont{D.}~\bibnamefont{Poulin}},
  \bibinfo{author}{\bibfnamefont{R.}~\bibnamefont{Blume-Kohout}},
  \bibinfo{author}{\bibfnamefont{R.}~\bibnamefont{Laflamme}}, \bibnamefont{and}
  \bibinfo{author}{\bibfnamefont{H.}~\bibnamefont{Ollivier}},
  \bibinfo{journal}{Physical review letters} \textbf{\bibinfo{volume}{92}},
  \bibinfo{pages}{177906} (\bibinfo{year}{2004}).

\bibitem[{\citenamefont{Shor and Jordan}(2007)}]{shor2007estimating}
\bibinfo{author}{\bibfnamefont{P.~W.} \bibnamefont{Shor}} \bibnamefont{and}
  \bibinfo{author}{\bibfnamefont{S.~P.} \bibnamefont{Jordan}},
  \bibinfo{journal}{arXiv preprint arXiv:0707.2831}  (\bibinfo{year}{2007}).

\bibitem[{\citenamefont{Boixo and Somma}(2008)}]{boixo2008parameter}
\bibinfo{author}{\bibfnamefont{S.}~\bibnamefont{Boixo}} \bibnamefont{and}
  \bibinfo{author}{\bibfnamefont{R.~D.} \bibnamefont{Somma}},
  \bibinfo{journal}{Physical Review A} \textbf{\bibinfo{volume}{77}},
  \bibinfo{pages}{052320} (\bibinfo{year}{2008}).

\bibitem[{\citenamefont{Cable et~al.}(2016)\citenamefont{Cable, Gu, and
  Modi}}]{cable2016power}
\bibinfo{author}{\bibfnamefont{H.}~\bibnamefont{Cable}},
  \bibinfo{author}{\bibfnamefont{M.}~\bibnamefont{Gu}}, \bibnamefont{and}
  \bibinfo{author}{\bibfnamefont{K.}~\bibnamefont{Modi}},
  \bibinfo{journal}{Physical Review A} \textbf{\bibinfo{volume}{93}},
  \bibinfo{pages}{040304} (\bibinfo{year}{2016}).

\bibitem[{\citenamefont{Lanyon et~al.}(2008)\citenamefont{Lanyon, Barbieri,
  Almeida, and White}}]{lanyon2008experimental}
\bibinfo{author}{\bibfnamefont{B.}~\bibnamefont{Lanyon}},
  \bibinfo{author}{\bibfnamefont{M.}~\bibnamefont{Barbieri}},
  \bibinfo{author}{\bibfnamefont{M.}~\bibnamefont{Almeida}}, \bibnamefont{and}
  \bibinfo{author}{\bibfnamefont{A.}~\bibnamefont{White}},
  \bibinfo{journal}{Physical review letters} \textbf{\bibinfo{volume}{101}},
  \bibinfo{pages}{200501} (\bibinfo{year}{2008}).

\bibitem[{\citenamefont{Hor-Meyll et~al.}(2015)\citenamefont{Hor-Meyll, Tasca,
  Walborn, Ribeiro, Santos, and Duzzioni}}]{hor2015deterministic}
\bibinfo{author}{\bibfnamefont{M.}~\bibnamefont{Hor-Meyll}},
  \bibinfo{author}{\bibfnamefont{D.}~\bibnamefont{Tasca}},
  \bibinfo{author}{\bibfnamefont{S.}~\bibnamefont{Walborn}},
  \bibinfo{author}{\bibfnamefont{P.~S.} \bibnamefont{Ribeiro}},
  \bibinfo{author}{\bibfnamefont{M.}~\bibnamefont{Santos}}, \bibnamefont{and}
  \bibinfo{author}{\bibfnamefont{E.}~\bibnamefont{Duzzioni}},
  \bibinfo{journal}{Physical Review A} \textbf{\bibinfo{volume}{92}},
  \bibinfo{pages}{012337} (\bibinfo{year}{2015}).

\bibitem[{\citenamefont{Passante et~al.}(2009)\citenamefont{Passante, Moussa,
  Ryan, and Laflamme}}]{passante2009experimental}
\bibinfo{author}{\bibfnamefont{G.}~\bibnamefont{Passante}},
  \bibinfo{author}{\bibfnamefont{O.}~\bibnamefont{Moussa}},
  \bibinfo{author}{\bibfnamefont{C.}~\bibnamefont{Ryan}}, \bibnamefont{and}
  \bibinfo{author}{\bibfnamefont{R.}~\bibnamefont{Laflamme}},
  \bibinfo{journal}{Physical review letters} \textbf{\bibinfo{volume}{103}},
  \bibinfo{pages}{250501} (\bibinfo{year}{2009}).

\bibitem[{\citenamefont{Passante et~al.}(2011)\citenamefont{Passante, Moussa,
  Trottier, and Laflamme}}]{passante2011experimental}
\bibinfo{author}{\bibfnamefont{G.}~\bibnamefont{Passante}},
  \bibinfo{author}{\bibfnamefont{O.}~\bibnamefont{Moussa}},
  \bibinfo{author}{\bibfnamefont{D.}~\bibnamefont{Trottier}}, \bibnamefont{and}
  \bibinfo{author}{\bibfnamefont{R.}~\bibnamefont{Laflamme}},
  \bibinfo{journal}{Physical Review A} \textbf{\bibinfo{volume}{84}},
  \bibinfo{pages}{044302} (\bibinfo{year}{2011}).

\bibitem[{\citenamefont{Wang et~al.}(2019)\citenamefont{Wang, Han, Yadin, Ma,
  Ma, Cai, Xu, Hu, Wang, Song et~al.}}]{wang2019witnessing}
\bibinfo{author}{\bibfnamefont{W.}~\bibnamefont{Wang}},
  \bibinfo{author}{\bibfnamefont{J.}~\bibnamefont{Han}},
  \bibinfo{author}{\bibfnamefont{B.}~\bibnamefont{Yadin}},
  \bibinfo{author}{\bibfnamefont{Y.}~\bibnamefont{Ma}},
  \bibinfo{author}{\bibfnamefont{J.}~\bibnamefont{Ma}},
  \bibinfo{author}{\bibfnamefont{W.}~\bibnamefont{Cai}},
  \bibinfo{author}{\bibfnamefont{Y.}~\bibnamefont{Xu}},
  \bibinfo{author}{\bibfnamefont{L.}~\bibnamefont{Hu}},
  \bibinfo{author}{\bibfnamefont{H.}~\bibnamefont{Wang}},
  \bibinfo{author}{\bibfnamefont{Y.}~\bibnamefont{Song}}, \bibnamefont{et~al.},
  \bibinfo{journal}{Physical Review Letters} \textbf{\bibinfo{volume}{123}},
  \bibinfo{pages}{220501} (\bibinfo{year}{2019}).

\bibitem[{\citenamefont{Krzyzanowska et~al.}(2017)\citenamefont{Krzyzanowska,
  Copley-May, Romain, MacCormick, and Bergamini}}]{krzyzanowska2017quantum}
\bibinfo{author}{\bibfnamefont{K.}~\bibnamefont{Krzyzanowska}},
  \bibinfo{author}{\bibfnamefont{M.}~\bibnamefont{Copley-May}},
  \bibinfo{author}{\bibfnamefont{R.}~\bibnamefont{Romain}},
  \bibinfo{author}{\bibfnamefont{C.}~\bibnamefont{MacCormick}},
  \bibnamefont{and}
  \bibinfo{author}{\bibfnamefont{S.}~\bibnamefont{Bergamini}}, in
  \emph{\bibinfo{booktitle}{Journal of Physics: Conference Series}}
  (\bibinfo{organization}{IOP Publishing}, \bibinfo{year}{2017}), vol.
  \bibinfo{volume}{793}, p. \bibinfo{pages}{012015}.

\bibitem[{\citenamefont{Fano}(1983)}]{fano1983pairs}
\bibinfo{author}{\bibfnamefont{U.}~\bibnamefont{Fano}},
  \bibinfo{journal}{Reviews of Modern Physics} \textbf{\bibinfo{volume}{55}},
  \bibinfo{pages}{855} (\bibinfo{year}{1983}).

\bibitem[{\citenamefont{Horodecki et~al.}(1995)\citenamefont{Horodecki,
  Horodecki, and Horodecki}}]{horodecki1995violating}
\bibinfo{author}{\bibfnamefont{R.}~\bibnamefont{Horodecki}},
  \bibinfo{author}{\bibfnamefont{P.}~\bibnamefont{Horodecki}},
  \bibnamefont{and}
  \bibinfo{author}{\bibfnamefont{M.}~\bibnamefont{Horodecki}},
  \bibinfo{journal}{Physics Letters A} \textbf{\bibinfo{volume}{200}},
  \bibinfo{pages}{340} (\bibinfo{year}{1995}).

\bibitem[{\citenamefont{Vidal and Werner}(2002)}]{vidal2002computable}
\bibinfo{author}{\bibfnamefont{G.}~\bibnamefont{Vidal}} \bibnamefont{and}
  \bibinfo{author}{\bibfnamefont{R.~F.} \bibnamefont{Werner}},
  \bibinfo{journal}{Physical Review A} \textbf{\bibinfo{volume}{65}},
  \bibinfo{pages}{032314} (\bibinfo{year}{2002}).

\bibitem[{\citenamefont{Baumgratz et~al.}(2014)\citenamefont{Baumgratz, Cramer,
  and Plenio}}]{baumgratz2014quantifying}
\bibinfo{author}{\bibfnamefont{T.}~\bibnamefont{Baumgratz}},
  \bibinfo{author}{\bibfnamefont{M.}~\bibnamefont{Cramer}}, \bibnamefont{and}
  \bibinfo{author}{\bibfnamefont{M.~B.} \bibnamefont{Plenio}},
  \bibinfo{journal}{Physical review letters} \textbf{\bibinfo{volume}{113}},
  \bibinfo{pages}{140401} (\bibinfo{year}{2014}).

\bibitem[{\citenamefont{Yu et~al.}(2016)\citenamefont{Yu, Zhang, Xu, and
  Tong}}]{yu2016alternative}
\bibinfo{author}{\bibfnamefont{X.-D.} \bibnamefont{Yu}},
  \bibinfo{author}{\bibfnamefont{D.-J.} \bibnamefont{Zhang}},
  \bibinfo{author}{\bibfnamefont{G.}~\bibnamefont{Xu}}, \bibnamefont{and}
  \bibinfo{author}{\bibfnamefont{D.}~\bibnamefont{Tong}},
  \bibinfo{journal}{Physical Review A} \textbf{\bibinfo{volume}{94}},
  \bibinfo{pages}{060302} (\bibinfo{year}{2016}).

\bibitem[{\citenamefont{{\.Z}yczkowski
  et~al.}(2011)\citenamefont{{\.Z}yczkowski, Penson, Nechita, and
  Collins}}]{zyczkowski2011generating}
\bibinfo{author}{\bibfnamefont{K.}~\bibnamefont{{\.Z}yczkowski}},
  \bibinfo{author}{\bibfnamefont{K.~A.} \bibnamefont{Penson}},
  \bibinfo{author}{\bibfnamefont{I.}~\bibnamefont{Nechita}}, \bibnamefont{and}
  \bibinfo{author}{\bibfnamefont{B.}~\bibnamefont{Collins}},
  \bibinfo{journal}{Journal of Mathematical Physics}
  \textbf{\bibinfo{volume}{52}}, \bibinfo{pages}{062201}
  (\bibinfo{year}{2011}).

\bibitem[{\citenamefont{Zyczkowski and Kus}(1994)}]{zyczkowski1994random}
\bibinfo{author}{\bibfnamefont{K.}~\bibnamefont{Zyczkowski}} \bibnamefont{and}
  \bibinfo{author}{\bibfnamefont{M.}~\bibnamefont{Kus}},
  \bibinfo{journal}{Journal of Physics A: Mathematical and General}
  \textbf{\bibinfo{volume}{27}}, \bibinfo{pages}{4235} (\bibinfo{year}{1994}).

\bibitem[{\citenamefont{Ozols}(2009)}]{ozols2009generate}
\bibinfo{author}{\bibfnamefont{M.}~\bibnamefont{Ozols}},
  \emph{\bibinfo{title}{How to generate a random unitary matrix}}
  (\bibinfo{year}{2009}).

\bibitem[{\citenamefont{Kent et~al.}(1999)\citenamefont{Kent, Linden, and
  Massar}}]{kent1999optimal}
\bibinfo{author}{\bibfnamefont{A.}~\bibnamefont{Kent}},
  \bibinfo{author}{\bibfnamefont{N.}~\bibnamefont{Linden}}, \bibnamefont{and}
  \bibinfo{author}{\bibfnamefont{S.}~\bibnamefont{Massar}},
  \bibinfo{journal}{Physical review letters} \textbf{\bibinfo{volume}{83}},
  \bibinfo{pages}{2656} (\bibinfo{year}{1999}).

\bibitem[{\citenamefont{Piani et~al.}(2011)\citenamefont{Piani, Gharibian,
  Adesso, Calsamiglia, Horodecki, and Winter}}]{piani2011all}
\bibinfo{author}{\bibfnamefont{M.}~\bibnamefont{Piani}},
  \bibinfo{author}{\bibfnamefont{S.}~\bibnamefont{Gharibian}},
  \bibinfo{author}{\bibfnamefont{G.}~\bibnamefont{Adesso}},
  \bibinfo{author}{\bibfnamefont{J.}~\bibnamefont{Calsamiglia}},
  \bibinfo{author}{\bibfnamefont{P.}~\bibnamefont{Horodecki}},
  \bibnamefont{and} \bibinfo{author}{\bibfnamefont{A.}~\bibnamefont{Winter}},
  \bibinfo{journal}{Physical review letters} \textbf{\bibinfo{volume}{106}},
  \bibinfo{pages}{220403} (\bibinfo{year}{2011}).

\bibitem[{\citenamefont{{\.Z}yczkowski and
  Sommers}(2005)}]{zyczkowski2005average}
\bibinfo{author}{\bibfnamefont{K.}~\bibnamefont{{\.Z}yczkowski}}
  \bibnamefont{and} \bibinfo{author}{\bibfnamefont{H.-J.}
  \bibnamefont{Sommers}}, \bibinfo{journal}{Physical Review A}
  \textbf{\bibinfo{volume}{71}}, \bibinfo{pages}{032313}
  (\bibinfo{year}{2005}).

\bibitem[{\citenamefont{Wootters}(1981)}]{wootters1981statistical}
\bibinfo{author}{\bibfnamefont{W.~K.} \bibnamefont{Wootters}},
  \bibinfo{journal}{Physical Review D} \textbf{\bibinfo{volume}{23}},
  \bibinfo{pages}{357} (\bibinfo{year}{1981}).

\bibitem[{\citenamefont{Debarba et~al.}(2012)\citenamefont{Debarba, Maciel, and
  Vianna}}]{debarba2012witnessed}
\bibinfo{author}{\bibfnamefont{T.}~\bibnamefont{Debarba}},
  \bibinfo{author}{\bibfnamefont{T.~O.} \bibnamefont{Maciel}},
  \bibnamefont{and} \bibinfo{author}{\bibfnamefont{R.~O.}
  \bibnamefont{Vianna}}, \bibinfo{journal}{Physical Review A}
  \textbf{\bibinfo{volume}{86}}, \bibinfo{pages}{024302}
  (\bibinfo{year}{2012}).

\end{thebibliography}

\end{document}